\documentclass[11pt]{article}
\def\beq{\begin{equation}}
\def\eeq{\end{equation}}
\def\bea{\begin{eqnarray}}
\def\eea{\end{eqnarray}}
 
\def\pa{\partial}

 \def\G{\Gamma}
\def\a{\alpha}
\def\b{\beta}
\def\d{\delta} 
\def\ve{\varepsilon}

\def\m{\mu}
\def\n{\nu}

\def\mn{{\mu\nu}}

\def\ij{{ij}}
\def\bma{\mbox{\boldmath$A$}}
\def\bme{\mbox{\boldmath$E$}}
\def\bmd{\mbox{\boldmath$D$}}
\def\bmn{\mbox{\boldmath$\nabla$}}
\def\bmb{\mbox{\boldmath$B$}}
\def\bmp{\mbox{\boldmath$P$}}
\def\bmv{\mbox{\boldmath$V$}}
\def\bmf{\mbox{\boldmath$F$}}
\def\half{{\textstyle{\frac{1}{2}}}}
\def\co{{\cal O}}
\def\be{\begin{equation}}
\def\bea{\begin{eqnarray}}
\def\ee{\end{equation}}
\def\eea{\end{eqnarray}}

\begin{document}

\begin{flushright}
BRX TH-560 \end{flushright}

\begin{center}
{\large\bf Free Spin 2 Duality Invariance Cannot be Extended to
GR}

{\large S.\ Deser$^1$ and D.\ Seminara$^2$}

{\it $^1$Department of Physics\\ Brandeis University\\
Waltham, MA 02454, USA\\{\tt deser@brandeis.edu}

$^2$Dipartimento di Fisica, Polo Scientifico \\ Universit\`{a} di
Firenze \\ INFN Sezione di Firenze Via G. Sansone 1  \\50019 Sesto
Fiorentino, Italy\\ {\tt seminara@fi.infn.it}}

\end{center}

\noindent {\bf Abstract}~ We show by explicit computation that the
recently discovered duality invariance of D=4 linearized gravity
fails, already at first self-interacting, cubic, approximation of
GR.  In contrast, the cubic Yang--Mills correction to Maxwell does
admit a simple deformed duality.

\section{Introduction}
\setcounter{equation}{0}
\def\theequation{\thesection.\arabic{equation}}

``Duality" has become a touchstone in relating seemingly different
regimes and models in field and string theory.  Its humble origins
lie in the ancient observation that in (and only in) D=4, on-shell
configurations with mutually rotated electric-magnetic fields also
obey the source-free Maxwell equations.  While this observation is
essentially correct, it is sometimes misunderstood.  Duality can
only be properly formulated in terms of the unconstrained
dynamical variables, rather than as a formal $\bme \leftrightarrow
\bmb$ rotation \cite{D+T}. This is already apparent from the facts
that $\bmb$ is identically divergenceless, and so can only be
related to $\bme$ upon implementing the latter's Gauss constraint,
and from the Maxwell Lagrangian's second order, ``hyperbolic",
non-dual invariant, form $L_M = \frac{1}{2} (\bme^2 - \bmb^2)$.

Surprisingly, duality invariance was recently \cite{H+T} extended
to free massless spin 2, then to free gauge fields of any spin and
statistics \cite{D+S}. Can this abelian duality invariance be
promoted to encompass the two physical self-interacting
generalizations, vector (YM) and tensor (GR)? [No consistent
nonabelian higher spin models are known.] The YM conjecture was
already considered, and settled in the negative in \cite{D+T},
although we will see that, actually, its first-cubic-extension
does allow a natural, consistent deformation of duality, at least
in Coulomb gauge. Our other, and principal, objective is to settle
the question (raised in \cite{H+T}) of extending duality to GR. We
will show (somewhat laboriously) that this cannot be achieved, at
least perturbatively: (deformed) duality already fails at its
first, interesting, cubic, level.

We will begin with a brief summary of what linear duality is (as
well as of what it is not) for free gauge fields: the two degrees
of freedom of all D=4 free gauge fields--their $\pm s$
helicities--can be rotated into one another, by a canonical
transformation mixing their two pairs of unconstrained dynamical
variables, while keeping the Hamiltonian form-invariant.  This is
not to be confused with other transformations, such as the
``harmonic oscillator" $(p \rightarrow q$, $q \rightarrow -p)$
rotations within a given mode.  We will then revisit the YM system
and establish its cubic order duality invariance, before turning
to the GR case.  An Appendix provides details of the latter.

\section{Free Vector Field Duality}
\setcounter{equation}{0}
\def\theequation{\thesection.\arabic{equation}}

Free gauge field duality is a canonical transformation, linking
coordinates and momenta of different excitations, that leaves the
Hamiltonian form-invariant.  The simplest case is Maxwell's, whose
first-order action is
 \be
 I_{max} [\bme^T\!,\bma^T] \!\! = \!\! \int \!\! d^4x \!
 \left[ (-\bme^T)\! \cdot \!\bma^T \! \! -\!\!
 \half (\bme^2_T \!+
\! (\bmn\!\times\! \bma^T)^2)\right] \!\!\equiv \!\! \int \!\!
d^4x \! \left[
 \sum^2_{A=1} p_A\dot{q}^A \! -\!  H (p_A, q^A )  \right]\! ,
 \ee
 upon implementing the Gauss constraint, $\bmn\cdot \bme=0$.
Here, transverse vectors are labelled with a ``$T$", from the
usual
 transverse-longitudinal decomposition,
 \be
\bmv =\bmv^T \! + \! \bmv^L \; , \;\;\bmn\cdot \bmv^T \equiv 0
\equiv \nabla \!\times\! \bmv^L \; , \;\; \int d^3x \: \bmv^T
\!\cdot \! \mbox{\boldmath$W$}^L = 0 \; .
 \ee
The indicated orthogonality between any pair of $T$ and $L$
vectors implies that only the manifestly gauge-invariant $\bma^T$
component survives in the action (2.1).  The duality rotation's
infinitesimal form is (we exhibit the -- redundant -- $T$-index on
$\bmb$ for emphasis):
 \be
 \d\bme^T =\bmb^T \equiv \bmn\! \times\! \bma^T\; , \;\;
 \d\bma^T =  \nabla^{-2} \bmn\! \times\! \bme^T \Rightarrow \d\bmb^T = -\bme^T \;
 .
 \ee
This $\bme^T \leftrightarrow \bmb^T$ rotation clearly leaves $\int
(\bme^2_T + \bmb^2_T)$ invariant. The middle equation demonstrates
that the desired $\bmb$-rotation is indeed implementable at the
level of the canonical coordinates $\bma^T$.  That (2.3) is also
canonical, {\it i.e.}, that the symplectic form $S \equiv \int
p\dot{q}$ term is invariant, follows from the fact that the curl
and Laplacian are hermitian operators, ${\cal O}$, by virtue of
which any $\int \chi {\cal O} \dot{\chi} = - \int \dot{\chi} {\cal
O}\chi = - \int \chi {\cal O}\dot{\chi} = 0$ upon (double) parts
integration. The loss of manifest Lorentz invariance and of space
locality inherent in this procedure is entirely harmless and
indeed necessary even to formulate, let alone establish the
transformations as canonical ones. [Note that it is only in three
space dimensions that the vector $\bme$ can even be matched with
the magnetic tensor $F_{ij}$ (by dualization in the $\ve^{ijk}$
sense).]

The above rotation is quite different from ``harmonic oscillator"
duality, valid in any dimension
 $$
 I [m,k]= \int [p\dot{q} - \textstyle{\frac{1}{2}} (p^2/m + kq^2)] \; ,
 $$
\be
\d p = \a q \; , \;\;\; \d q = - \b p \; , \;\;\; \a /\b = km \; .
\ee
 that relates a single excitation's variables, and reflects the
equivalence of different parameter regimes\footnote{Or relating
different models, here $I[m,k] \leftrightarrow I [k^{-1},
m^{-1}]$, with different parameters \cite{zwiebach}.} through the
dependence of the solutions on $(k/m)$. There is a similar
``duality" of Maxwell theory: rewriting (2.1) as
 \be
 I_M = \int \left\{ (-\bme)\!\cdot\!
 \dot{\bma} - \textstyle{\frac{1}{2}}
 [\bme^2 +\bma (-\nabla^2)
\bma] \right\} \; ,
 \ee
immediately implies invariance, within each helicity, under
$$
\d\bme = \sqrt{-\nabla^2} \: \bma\; , \;\;\;\d\bma = - (-\nabla^2
)^{-1/2}\bme \; .
$$

\section{Cubic Yang--Mills Duality}
\setcounter{equation}{0}
\def\theequation{\thesection.\arabic{equation}}
As an instructive (and transparent) contrast to GR, we study first
the nonlinear extension of a Maxwell multiplet by adding the cubic
terms in YM, and show that, surprisingly, they permit--a deformed
version of--abelian duality.  [This in no way contradicts the
demonstration in \cite{D+T} that full YM precludes duality.]  For
simplicity, we work with SU$_2$, whose structure constants
$\ve^{abc}$ permit an obvious 3D internal vector notation. The
first-order covariant YM action is (setting $g=1$),
 \be
 I_{YM} [\bmf,\bma] \equiv -\half \int \left\{ \bmf^\mn
 \! \cdot \!
 (\pa_\m \bma_\n \!\!-\!\! \pa_\n \bma_\m \!\!+\!\! \bma_\m \!\!\times\!\! \bma_\n )
 -\half
 \bmf^\mn \!\cdot\! \bmf_\mn \right\} \; .
 \ee
The (3+1) versions of (3.1) and of the Gauss constraint become
\be
 I_{YM} [ \bme,\bma ] = \half
\int [ (-\bme)\! \cdot\!\dot{\bma} -
\textstyle{\frac{1}{2}}(\bme^2 + \bmb^2)] , \;\;\bmb \equiv \bmn
\times\bma + \bma \!\times \! \bma  ,\;\; \bmd\cdot \bme \equiv
(\bmn \!+ \!\bma \times) \cdot \bme = 0 .
 \ee
Adopting Coulomb gauge, $\bma^L = 0$ simplifies the process:
 \be
 I_{YM} =\half \int \left\{ (-\bme^T)
 \!\cdot\! \dot{\bma}_T -\half
 [\bme^2_T + \bme^2_L +
\bmb^2 ]\right\} \; ,
 \ee
 \be
 (\bmn + \bma_T \times ) \cdot
 \bme_L = - \bma_T \!\times\! \bme_T
 \; .
 \ee
The constraint\footnote{Formally, inversion of (3.4) for $\bme_L
[\bma_T, \; \bme_T]$ would solve classical YM entirely!} fixes
$\bme_L$ in terms of the dynamical pairs: $\bme_L = 0 + {\cal O}
(\bma^T \bme^T)$; since $\bme^2_L$ is quartic, we may drop it from
(3.3) to cubic order; omitting ``$T$", this leaves
 \be
 I_{YM} [\bme , \bma ]
 = I_M [\bme , \bma ] + I^c_{YM} [\bme , \bma ]
 \cong \half \!\int\! \left\{
 ( -\bme )\! \cdot\! \dot{\bma} \!\!-\!\!\half (\bme^2\!\! +\!\!
 (\bmn \! \times \! \bma )^2 \right\}  + \!\int\! \bmn \!\!\times\!\! \bma \!\cdot\!
  \bma \!\times\! \bma
 \; .
 \ee
 Only the final term in $H$
 differentiates the action from that of a triplet of
 photons.  The original linear duality transformation (2.3) clearly
 alters (only) this term,
 \be
 \d_L \int (\bmn\! \times\! \bma \!\cdot\! \bma \!\times\! \bma ) = 3\int \bme
 \!\cdot\! \bma \!\times\! \bma \neq 0 \; .
 \ee
 To cancel this cubic term, we must deform the abelian transformation by
 adding a quadratic $\d_Q
 \bme$, that will generate a cubic variation from the $\int
 \bme^2$ in (3.5); the obvious choice is
 \be
 \d_Q \bme = (\bma \!\times\! \bma )^T \; .
 \ee
[We have projected the ``$T$" part of $(\bma \!\times\! \bma )$
since $\bme$ is transverse; this is just a formality here, and
throughout, since orthogonality would automatically perform the
projection in $\int \bme^T\!\cdot\d_Q\bme$.]  Having succeeded in
keeping $H_{YM}$ invariant, we must still check that the
symplectic variation, generated by (3.7), namely
  \be
  \d_Q \int \bme \!\cdot\! \dot{\bma} = \int \bma \!\times\! \bma\! \cdot\! \dot{\bma}
 \ee
 vanishes.  Indeed, time integration by parts shows
 it to equal minus twice itself.

To summarize, we have succeeded in keeping duality invariance of
YM to lowest nonlinear order, in $\bma_L=0$ gauge at least. This
was accomplished by setting $\d\bme = \bmb \equiv \bmn \!\times\!
\bma + \bma\! \times\! \bma$, certainly the most obvious guess;
the corresponding part of the generator is just the YM
Chern--Simons form:
 \bea
 G_E &  = & \int d^3x [\bma \! \cdot \! \bmn \! \times \! \bma +
  {\textstyle\frac{2}{3}} \bma \! \cdot \! \bma \!\times\!\bma ] \; , \nonumber \\
 && \d \bme = [G_E,\bme ] = \bmn \!\times\! \bma\! + \!\bma \!\times\! \bma
 \equiv \bmb_{YM}
 \; .
 \eea
 We have not attempted to extend this process to quartic order,
where terms $\sim (\bma\! \times\! \bma )^2$ from $\bmb^2$ and
$\sim (\bma \cdot \bme_T)^2$ from $\bme^2_L$ appear, nor do we
know whether the cubic order success has some deeper physical
origin.

 \section{GR Duality Fails}
\setcounter{equation}{0}
\def\theequation{\thesection.\arabic{equation}}
 Having illustrated how duality deformations can succeed at cubic
level for YM vectors, we now turn to the--considerably more
complicated--tensor case.  We will begin with a description of
Pauli--Fierz free spin 2 theory, transcribed into a notation
manifesting its duality invariance.
 We will then derive the cubic correction to GR, and subject it to
 the abelian transformations.  Finally, we will show that its lack
 of invariance under the latter cannot be compensated by adding
 further, quadratic, deformations of the dynamical variables:
 there is no (perturbative) duality invariance in GR.

Let us first express the familiar free spin 2 gauge system in
``dual-ready" form, using the first order formulation of full GR
\cite{ADM}, thereby also obtaining the cubic correction in a
unified way.  The Einstein action is
 \bea
 I_E [\pi , g] & = & \int d^4x \; [\pi^{ij} \dot{g}_{ij} - N_\m R^\m
 ] \nonumber \\
 && N_0 \equiv (-g^{00})^{-1/2} \; , \;\;\; N_i \equiv g_{0i}
 \nonumber \\
 R^0 & \equiv & -\sqrt{^3g}\;\, ^3\!R + (\pi_{ij}\pi^{ij} -
 \textstyle{\frac{1}{2}} \pi^i_i \pi^j_j )(^3\!g)^{-1/2} \; , \;\;\;
 R^i \equiv -2 D_j\pi^{ij} \; .
 \eea
 All explicit (roman) indices refer to the intrinsic 3-space, of
 which $^3R$ is the scalar curvature and $D_j$ the covariant 3-derivative;
$\pi^{ij}$, essentially the second fundamental form (density), is
an independent variable on a par with $g_{ij}$. Our convention is
$R_{ij} \sim + \pa_k \G^k_{ij}$. The expansion about flat space
is\footnote{We spell out the fact that $g_{ij}$ is exactly
$\d_{ij} + h_{ij}$ to all orders, so that neither $h_{ij}$ nor
$\pi^{ij}$ are subject to further expansion, unlike say $g^{ij}
\sim \d_{ij} - h^{ij} + {\cal O}(h^2)$ or $\sqrt{^3g} \sim 1 +
\half \, h_{ii} + {\cal O} (h^2)$.}
 \be
 h_{ij} \equiv g_{ij} - \d_{ij} \; , \;\;\; N_0 = 1+n \; ;
 \ee
 instead, $N_i$ and $\pi^{ij}$ vanish in flat space and are of
first order. Abelian gauge invariance and use of the four
constraints, $R^\m = 0$ simplifies (4.1) dramatically.  We recall
the familiar transverse-traceless orthogonal decomposition of a
flat space (where index position is immaterial) spatial tensor
$T_{ij} = T_{ji}$,
 \bea
 T_{ij} & = & T^{TT}_{ij} + \textstyle{\frac{1}{2}} (\d_{ij} -
 \nabla^{-2} \pa^2_{ij} ) T^T + (\pa_i T_j + \pa_j T_i ) \nonumber
 \\
 \pa_i T^{TT}_{ij} & = & 0 \; =\; T^{TT}_{ii}\; , \;\;\;\; \pa_j
(\d_{ij} - \nabla^{-2} \pa^2_{ij} ) \equiv 0 \equiv (\d_{ij} -
\nabla^2 \pa^2_{ij} ) \pa_j  \;  .
 \eea
 Of the six components of $T_{ij}$, two are ``$TT$" one
is ``T" and three represent the vector $T_i$. Orthogonality
between the various components under integration is manifest; one
example is $\int d^3x \: T^{TT}_{ij} \pa_i W_j = 0$. Once
decomposed in this fashion, the twelve $(\pi^{ij} , h_{ij})$
components are easily classified: At linearized level, the
theory's abelian invariance under $\d h_\mn = (\pa_\m \xi_\n +
\pa_\n \xi_\m )$, simply means that the four gauge variables
$(\pi^T, h_i)$ do not appear in the quadratic action. Likewise,
the four linearized constraints,
 \be
 -R^0 \rightarrow \; ^3\!\!R_L (h) \equiv (\pa^2_{ij} h_{ij} -
  \nabla^2 h_{ii}) = - \nabla^2 h^T = 0 \; ,
   \;\;\; R^i \rightarrow -2 \pa_j \pi^{ij} = 0
 \ee
 remove the respective components $(h^T, \pi^i)$, as is also obvious for
 $\pi^i$ upon using (4.3).
The four remaining $(\pi^{ij}, h_{ij})$ then simply reduce to
their two pairs of ``$TT$" degrees of freedom.  We designate them
by\footnote{More precisely, the original linear Hamiltonian is
$\int [\pi^2_{ij} + \frac{1}{4} (\bmn h_{ij} )^2]$, requiring the
rescaling $\sqrt{2} \, \pi^{ij} = p^{ij}$, $1/\sqrt{2} \, h_{ij} =
q_{ij}$ to bring it into standard free-field form (4.6).}
$(p^{ij}, h_{ij})$ henceforth dropping the ``$TT$" notation. The
symplectic term is just that of the two excitations,
  \be
S = \int d^4x \: \pi^{ij} \dot{g}_{ij} \rightarrow \int d^4x \:
p^{ij} \dot{q}_{ij}\; ,
 \ee
 while the Hamiltonian is simply the quadratic part of $R^0$,
 subject to the linear constraints.  A short calculation
 (essentially integrations), finally reduces (4.1) to the usual
 harmonic oscillator form,
 \be
 I_E^Q [p,q] = \int d^4x \left\{ \sum^2_{A=1} p_A\dot{q}^A
 -
 \half [p^2 + (\bmn q)^2] \right\} \; ,
 \ee
 with traces on implicit indices understood.
For our purposes, the notion of duality in this symmetric tensor
world emerges from the existence of a generalized curl \cite{D+S},
distributed on the tensor's indices:
 $$
 (\co T)_{ij} \equiv \half (\ve^{i\ell m} \pa_\ell T_{mj} +
 \ve^{j\ell m} \pa_\ell T_{im} ) = (\co T)_{ji} \; . \eqno{(4.7{\rm
 a})}
 $$
Its nonlocal extension
$$
 (\co T)_{ij} \equiv (\nabla^{-2} \co T)_{ij} \eqno{(4.7{\rm
 b})}
 $$
 will also streamline notation; both $\co$-operations are
 hemitian.  Thus, acting on $q_{ij}$ the curl defines a
 ``magnetic" field,
\setcounter{equation}{7}
\def\theequation{\thesection.\arabic{equation}}
 \be
 B_{ij} \equiv  (\co q )_{ij} = B_{ji} \; ,
 \ee
  in terms of which (4.6) becomes
  \be
  I^Q_E [p,q] = \int d^4x \: \{ p\dot{q}  - \half (p^2 +
  B^2 ) \} \; .
  \ee
This form makes it irresistible to perform the
rotation\footnote{Higher free spin fields \cite{D+S} follow the
same pattern, in terms of suitably generalized curl operations to
accommodate any number of spatial indices.}
 \be
 \d p_{ij} = B_{ij} \; , \;\;\;
 \d q_{ij} =  (\tilde{\co}p)_{ij} \Rightarrow
 \d B_{ij} = - p_{ij} \; ,
 \ee
 the final step uses $\co\co = - \nabla^2$, just as
 $\nabla\times\nabla\times = - \nabla^2$ on transverse vectors.
This transformation obviously maintains the Hamiltonian of (4.6),
while the $q$-sector of (4.10) shows that $\d B$ is indeed
implementable as a transformation of the underlying coordinate
$q$.  Invariance of the symplectic term is guaranteed (as for spin
1) by hermiticity of the $\tilde{\co}$ and $\co$ operators.

Establishing duality transformation and invariance of free spin 2
under them was the easy part; now we must return to the full
action (4.1) to obtain its cubic, $I^c_E [p,q]$, correction,
subject it to the linear rotations (4.10), and try to cancel away
their (nonvanishing) effect by quadratically deforming them.

The cubic correction to $H$ is formally easy to find.  It is
simply the cubic part, $R_c^0$, subject to the lower level
constraints. The procedure is greatly simplified, as it was for
YM, by choosing ``Coulomb" gauge, here $h_i \!=\! 0\! =\! \pi^T$.
[Recall that only the linearized action is invariant under the
linear part of the non-abelian $\d g_\mn = D_\m \xi_\n + D_\n
\xi_\m$.]  The upshot of this process is the cubic Hamiltonian
 \be
 H_c (p,q) = R_c^0 (p,q) = \int d^3x \left[ (-^3\!R\sqrt{g}\,)_c + 2p^{ij}
 p^{\ell m} (q_{i\ell} \d_{jm} - \half q_{ij} \d_{\ell m} )
 \right] \; ;
 \ee
the last term in (4.11) vanishes due to $p^{\ell m}\d_{\ell m}=0$.
Furthermore, the symplectic term has the great virtue of remaining
quadratic in our gauge, while the lower constraints also imply
that the same $(p,q)$ are dynamical.

Whereas the cubic order curvature scalar density is a bit
elaborate, we are in fact only interested in its (linear)
variation;
 \be
 \d_L \int \: (^3\!R\sqrt{g}\,)_c = - \int (\sqrt{g} \: G^{ij})_Q \: \d_L q_{ij}
 \; ,
 \ee
 by the Palatini identity.  The variation of the cubic kinetic
 terms is just  $\d_L \int 2trppq$, so we have
 \be
 \d_L H_c (p,q) = \int d^3r\: G^Q_{ij} \d_L p^{ij} + 2 \int
 (pp)_{ij} \d_Lq_{ij} + 2 \int \{  \d_L p^{ij}(pq)_{ij} +
 (p\d_L p)_{ij}q_{ij} \}\; .
 \ee

Schematically, the variation (4.13) is of the form $\d_L I \sim
q^2p + p^3$.  There is an {\it a priori} daunting array of
possible compensating quadratic deformations of each $(p,q)$: $
\d_Q (p,q) \sim (p^2 + pq + q^2)$, with various possible
derivative and index structures in each.  Further, all are subject
to keeping the relative dimensions of $(p\sim \dot{q},q)$ as well
as enforcing $\d\int p\dot{q} = 0$. Owing to the dimensional
asymmetry between $\d_Qp$ and $\d_Qq$, the latter will involve the
non-local operator $\tilde{\co}$, as compared to $\d_Qp\sim \co$.
This will lead to the presence of both local and nonlocal
variations that must separately cancel.

 [Parenthetically, we lay to rest the otherwise attractive idea
 that, since duality invariance requires the other Poincar\'{e}
 generators of the cubic model to be invariant too, why not test the
 momentum generator instead, say?  Unfortunately, Dirac's dictum
 that the Hamiltonian form of dynamics always has a simple
 momentum holds here as well.  The transformation generator $G =
 P_i \d x^i$, with $P_i = -2 \int \pa_j \pi_i^j$ and the
 constraint $D_j \pi^{ij} = 0$ says that to all orders,
 \be
 \pa_j \pi_i\!\!~^j = \G_j\!\!~^k\!\!~_i \: \pi^j_k = [j i, k]
 \pi^{jk} = \half p \nabla_i q \; .
  \ee
Hence $\bmp = \int p \bmn q$ to all orders \cite{ADM} and its
invariance provides no independent test beyond that of $\int p
\frac{\pa}{\pa t} q$.  The same holds for the rotation generator,
of course.]

In the following, we will just outline the flow of possible
deformations and their consequences.  The more unpleasant details
are relegated to the Appendix. Returning to (4.13), we have
 \bea
 \d_L H_c (p,q) & = & \int d^3r \left[ G^{ij}_Q \: (\tilde{\co}p)_{ij}
 + 2 tr \: pp \tilde{\co}p + 2 tr
 \: \{ qp {\cal O}q + q {\cal O}q p \}\right] \nonumber \\
 & \sim & p^3 + q^2p \; ,
 \eea
 Obviously, the simplest term to cancel is the pure $p^3$, which
can (only) be accomplished by $\d^{(1)}_Q p \sim p^2$. More
precisely, acting on the quadratic Hamiltonian, this deformation
leads to\footnote{Here, and in all quadratic transformations, a
projection to ``$TT$" space is understood, to keep the character
of $(p,q)$. However, this is really immaterial, since they always
safely multiply a linear $TT$ variable.}
 \be
 \d^{(1)}_Q p_{ij} \sim (p \tilde{\co}p)_{ij} \Rightarrow
 \d^{(1)}_Q \int p^2 \sim tr (pp  \tilde{\co} p) \; ,
 \ee
 exactly of the form of the term to be cancelled.
 Now, however, we have to cancel the unwanted effect of (4.16) on $\int
 p\dot{q}$,
 \be
 \d^{(1)} \int p\dot{q} \sim tr \int (p \tilde{\co}p)_\ij
 \dot{q}_{ij} \; .
 \ee
 The unique possible cure for this is
\be
 \d^{(1)}_Q q \sim (q \tilde{\cal O}p+p\tilde{\cal
 O}q)_\ij \; ;
\ee
 and indeed it works, cancelling (4.17). While we have now gotten
 rid of the  $p^3$  part of
(4.15), we must not forget one further effect, that of (4.18) on
$H_Q$,
 \be
 \d^{(1)}_Q \half \int q_\ij (-\nabla^2)q_\ij \sim tr \int q
 (-\nabla^2) [q  \tilde{\cal O} p + p \tilde{\cal O}q]
 \sim q^2p \; ,
 \ee
a generically nonlocal term.  The possible remedies to the overall
$\int q^2p$ residue of (4.15) plus (4.19) consist of $\d^{(2)}_Q p
\sim q^2$ and $\d^{(2)}_Q q \sim qp$.  The latter choice is
clearly not desirable, since we have just exploited it in (4.18),
in an essentially unique way. We are therefore stuck with the last
hope,
 \be
 \d^{(2)}_Q p \sim q {\cal O} q \; , \;\;\;\;
 \d^{(2)}_Q  \half \int p^2 = tr \int p (q {\cal O} q) \; .
 \ee
Now we exploit the local/nonlocal division of the variations.  In
(4.15) there are both kinds; the former type is easy, since (by
dimensions) it has no explicit derivative beyond the curl $\co$,
that is,
\be
 \d_{local} H_c \sim 2 \int  \{ p({\cal O}q) q + q {\cal O}qp \}\;
 .
 \ee
 This is clearly cancelled by (and only by)
\be
 \d^{(2)}_Q p \sim 4(q
{\cal O}q)\; , \;\;\;\; \d \half \int p^2 =
 4 tr \int p({\cal O}q)q \; .
 \ee
 The symplectic contribution of (4.22) ``miraculously" vanishes;
 its (local) form is
\be
 \d^{(2)}_{local} \int p\dot{q} \sim tr \int(qq)_\ij ({\cal O}\dot{q})_\ij\; .
 \ee
The positions of the dot and the curl are immaterial.  What
matters is that $\co$ is hermitian, while $\pa /\pa t$ changes
sign on integration by parts, which suffices to show this term is
proportional to minus itself and vanishes.  The (decisive)
nonlocal terms require the Appendix!

 \section{Summary}
\setcounter{equation}{0}
\def\theequation{\thesection.\arabic{equation}}
 We have considered whether any deformations of the duality
transformations of linearized vector and tensor gauge theories
might rescue (an extended version of) duality in their nonlinear
regimes.\footnote{The vector and tensor models are of course very
different, not only in index complication, but in the fact that
cubic and higher YM terms are of lower derivative order, whereas
in GR, all order have two derivatives.}

  For YM, it was surprisingly easy, at least in Coulomb
gauge, to do so at leading, cubic, ``post-Maxwell" order, simply
by letting the  electric field variable rotate into the full YM
magnetic field, while keeping the vector potential's rotation
unaltered.  We have  not explicitly analyzed quartic order and
beyond in the theory's  infinite series expansion in $(p,q)$,
since it is known that duality fails for full YM.

Extension of free spin 2 duality to cubic order proved
considerably more complicated to decide, but we were able to show
that no deformation compensates for the loss of abelian duality
invariance, even in Coulomb gauge. We conclude that, perhaps
disappointingly, rotation among helicities ceases to be an
invariance beyond the free spin 2 level.

This work was supported in part by NSF grant PHY04-01667.

\appendix{}
\section{Appendix}
\setcounter{equation}{0}
\def\theequation{A.\arabic{equation}}
In section $4$, we sketched problems of extending linear spin 2
duality to the cubic term of GR.  This Appendix provides details
of the obstacles and shows why they cannot be overcome. We first
collect the key quantities, starting with the gravity Hamiltonian
at quadratic
 \beq \label{app1}
  H_Q =\half \int d^3 x~\left[ (\bmn q)^2 + p^2 \right],
   \eeq
 and cubic,
\be
\label{app2}
 H_c (p,q) = \int d^3x \left[ -^3 R_c (q) +
 2p^{i}_{\ m}
 p^{\ell m} q_{i\ell}
 \right]
 \ee
 levels, with all relevant constraints satisfied: only ``$TT$"
 variables are involved.
[The corrections coming from solving the nonlinear constraints
affect only quartic terms.] The abelian duality transformation
(4.10), which preserves $H_Q$, is
 \bea \label{app3}
  \delta^{(1)}
q_{ab}= (\tilde{\co} p)_{ab} \equiv P_{ab}
 \ \ \ \ \ \ \ &&\delta^{(1)} p^{ab}= (\co q)_{ab} \equiv Q_{ab}.
 \eea
 We now consider the effect of this rotation on $H_c$, starting with
the variation of the simple cubic vertex $ppq$ of (\ref{app2})
  \beq \label{app4}
-2 \d \int ppq =  -2 \int \delta^{(1)} q_{ab} p^{ac}p^{b}_{c}  + 2
q_{ab} \delta^{(1)} p^{ac} p^{b}_{c}  \equiv A + B.
 \eeq
The $A$ term's contribution in the above variation is proportional
to $p^3$. Specifically we have
 \be \label{app5}
  -2 \left(\delta^{(1)} q_{ab} \right) \pi^{ac}\pi^{b}_{c} =
  P_{ab} (pp)_{ab} \; .
  \ee
 The unique and obvious way to cancel the above variation
introduces a contribution in $\d p$ quadratic in momentum,
 \be  \label{app6}
\delta^{(2)}p_{bc}=- P_{ab} p_{bc} \; .
 \ee
 The ensuing variation of the  $p^2$ term in $H_Q$ then exactly cancels
(\ref{app5}). This deformation of $\d p$ obviously affects the
symplectic form $S$, which in turn must be compensated. Indeed,
 \be \label{app7}
 \d S = \int d^4x \: \d^{(2)} p^{ab}\dot{q}_{ab} = - \int d^4x
 P_{ab} p_{bc} \: \dot{q}_{ac} \; .
 \ee
Cancelling this variation requires, and completely  fixes, a
quadratic term $\sim pq$ in $\d q_{ab}$.  [Any alteration would
automatically produce a change in the $p^2$ term
$\delta^{(2)}p^{ab}$, which, however, has been just determined.]
To evaluate  the required $qp$ term in $\delta^{(2)}{q}_{ab}$, we
first integrate  (\ref{app7}) by parts with respect to   time,
 \be  \label{app8}
 \d  S = \int d^4x \: tr [q P\dot{p} +
 (pq)\dot{P} ] \; .
 \ee
  Thus, the desired variation must be
   \be \label{app9}
    \delta^{(2)}_{qp}q_{ac} = [q_{cb}P_{ba} + (ab)]
 \ee
Next we consider the $B$ term  in (\ref{app4}), namely
 \beq
\label{app10} -4\int d^3 x\ q_{ab}\left(\delta^{(1)}p^{ac}\right)
p^{b}_{c}= -4 \int d^3 x\ tr (qQp) \;.
 \eeq
  Variations of the
same $pq^2$ form will also contribute from $^3\!R_c (q)$ in
 (\ref{app2}),
  \be  \label{app11}
   -\delta\int d^3 x
^{(3)} R= -\int d^3 x ^{(2)} G^{ab}(q)\delta^{(1)} q_{ab}
 =2 \int
d^3 x  \: R^{ab}_Q \: P_{ab} \; ,
 \ee
 as well as from the $q^2$ term in $H_Q$
  \be  \label{app12}
-\half\int d^3 x~ q_{ab}\nabla^2 \d^{(2)}_{qp}q^{ab}=- \half \int
d^3x \: tr [q\nabla^2 q P + qQp ] \; .
 \ee
 The variation in (\ref{app11}) can be re-arranged by
explicitly displaying the Ricci tensor\footnote{We are allowed to
drop $(g^{ab}R)^Q = \d^{ab}R^Q - h_{ab}R_L$; the first of these
vanishes when contracted with a $TT$ tensor, the second because
$R_L =0$ is the linear constraint.}
 to $\co (q^2)$:
 \bea  \label{app13}
 ^{(2)}
R^{ab} &=& \left( \half\: q_{ac} \nabla^2 q^{c}_{b} +\half
\:q_{bc} \nabla^2 q^{c}_{a}\right) -\half\:\partial_r\left( q^{r
n}\left(\partial_a q_{nb}+
\partial_b q_{na}-\partial_n
q_{ab}\right)\right)+{\textstyle{\frac{1}{4}}}\:
\partial_a \partial_b \left(q^{rn}q_{rn}\right)\nonumber\\
&-& {\textstyle{\frac{1}{4}}}\:\partial_a q_{mr} \partial_b
q^{mr}-{\textstyle{\frac{1}{4}}} \left(\partial^r q^m_b-\partial^m
q^{r}_{b} \right)\left(\partial_m q_{ar}-\partial_r q_{am} \right)
\; .
 \eea
The $\d_a\d_b$  term can be dropped because it is longitudinal and
vanishes when integrating with a $TT$ object. The variation
(\ref{app11}) can be now rewritten as follows
 \bea \label{app14}
&&2 \int d^3 x  R^{ab}_Q(q)P_{ab} =\int d^3 x \biggl[ q_{ac}
\nabla^2 q^{c}_{b} + q_{bc} \nabla^2 q^{c}_{a} -\partial_l \bigl(
q^{l n} (\partial_a q_{nb}+
\partial_b q_{na}-\partial_n q_{ab})\bigr) \nonumber\\
&&- \half \: \partial_a q_{ml} \partial_b q^{ml}- \left(\partial_m
q_{al}\partial^l q_{bm} -  \partial_m q_{al} \partial^m
q^l_b\right) \biggr] P_{ab}\;.
 \eea
 The combination of the variation (\ref{app14}) and
(\ref{app12}) simplifies a bit to
 \bea \label{app15}
&&\int d^3 x  \biggl[\biggl\{\bigl( -\partial_l( q^{l n}
(\partial_a q_{nb}+
\partial_b q_{na}-\partial_n q_{ab}))-
\half\partial_a q_{ml} \partial_b q^{ml}-\left.
\partial_m q_{al}\partial^l q_{bm} \biggr\}P_{ab} \right.\nonumber\\
&&+ \half {(q_{ac} q^{c}_b)}P_{ab} -\half {q}^{m}_b Q_{ab}p_{am}
\biggr].
 \eea
  Finally, inclusion of (\ref{app10}) yields the complete cubic
variation,
 \bea \label{app16}
\d_{tot} H & = &\int d^3 x \biggl[\biggl\{\bigl( -\partial_l( q^{l
n} (\partial_a q_{nb}+
\partial_b q_{na}-\partial_n q_{ab}))-
\half\partial_a q_{ml} \partial_b q^{ml}-\left.
\partial_m q_{al}\partial^l q_{bm} \biggr\}P_{ab}\right.\nonumber\\
&+ &  \half \, tr (qqP) +\half {q}^{m}_b Q_{ab}p_{am} \biggr].
 \eea
 In order to cancel this variation,  as
already explained in text, we may only modify $\delta^{(2)}p$ by
adding to it terms proportional to $q^2$, while preserving the
symplectic form without help from other sectors. It produces, in
the variation of the symplectic form, a contribution of the type
$q^2\dot{q}$ that cannot originate from anywhere else.

In this remaining variation (\ref{app16}), we see two
``non-interacting" sectors: the first is non-local ($\nabla
^{-2}$) and the  second local; this splitting is unambiguous.
Inspecting the position of the indices of the derivatives in the
non-local part, one sees that they cannot produce further local
contribution by integration by parts. Hence the two contribution
must be cancelled by separate terms in the variation
$\delta^{(2)}p_{ab}$. Let us dispose first of the local sector.
Here, we modify $\delta^{(2)}p_{ab}$ as follows
 \beq
\delta^{(2)}_{q^2}p^{ab}= \co (qq)_{ab}  - q_{ac} Q_{cb} + (b
\leftrightarrow a) \; .
 \eeq
  The resulting variation of the symplectic form vanishes.
 \be
\d_{q{^2}} S = \int d^4 x  \left( \half\dot{q}_{ab}\epsilon^{m r
a}\partial_r{(q_{mc} q^{cb})} -\half\epsilon_{ \ \ }^{m r
a}\partial_r({q}_{mc}{q}^{cb}) {\dot{q}_{ab}}\right)=0.
 \ee
 As explained in text, any term $\sim \int (\co \dot{q})qq$
 effectively changes sign upon simultaneous space and time
 integrations by part.  This brings us down to the non-local part,
 the final (and as we now see, incurable) obstacles:
\be
\d_{NL} H_c = -\! \int \! d^3 x\biggl[\partial_l( q^{l n}
(\partial_a q_{nb}+
\partial_b q_{na}-\partial_n q_{ab}))+
\half\,\partial_a q_{ml} \partial_b q^{ml}+ ( \partial_m
q_{al}\partial^l q_{bm})\biggr] P_{ab}
 \ee
  This proliferation of
terms represents the additional index possibilities once the extra
$\nabla^{-2}$ brings in two extra derivatives in the numerator.

Write (A.19) schematically as
 \be
 \d_{NL} H_c = \int d^3x [qq] P \equiv \int d^3x p\tilde{\co}
 [pp] \; .
 \ee
As in the local term's analysis, (A.20) can be cancelled  by
$\d^{(2)}_{NL} p_{ab} = \{\tilde{\co} [ qq] \}_{ab}$ acting on the
$p^2$ part of $H_Q$, and the last hurdle is the ensuing symplectic
variation, engendered by $\d^{(2)}_{NL} p$:
 \be
 \d_{NL} S = \int qq\tilde{\co} \dot{q} \; ,
 \ee
 which must again vanish by itself, there being no further
 variations left to help it.  From (A.19), then,
 \be
 \d_{NL} S = \int d^4x \biggl\{ \pa_{\ell m} ( q_{a\ell}
 q_{bm} - q_{ab} q_{\ell m}) + \half \, q_{m\ell ,a} q_{m\ell ,
 b}  +  2\pa_\ell (q_{\ell m} q_{ma,b} ) \biggr\}
 \tilde{\co}\dot{q}_{ab}\; .
 \ee
The terms in curly brackets are essentially all the possible forms
involving $qq$ and $\pa\pa$ as a 2-index tensor: two, one and no
``dummy" derivatives.  Unlike in the local term's $qq\dot{q}$
there are no cancellations here and indeed there are concrete
counterexamples of $TT$ $q$-tensors with $\d_{NL} S \neq 0$.

\end{document}